\documentclass[aps,prl,twocolumn,superscriptaddress,preprintnumbers,%
               showpacs,nofootinbib]{revtex4}
\newcommand{\PRE}[1]{}       % Use if journal style
\usepackage{bm}
\usepackage{epsfig}

\newcommand{\postscript}[2]{\setlength{\epsfxsize}{#2\hsize}
   \centerline{\epsfbox{#1}}}

\def\to{\rightarrow}

\def\bi{\begin{itemize}}
\def\ei{\end{itemize}}

\def\tb{\tilde b}

\def\tst{\tilde t}
\def\ttau{\tilde \tau}

\def\tw{\widetilde W}
\def\tz{\widetilde Z}
\def\alt{\stackrel{<}{\sim}}

\def\be{\begin{equation}}  
\def\ee{\end{equation}}  

\begin{document}

\preprint{FSU-HEP-041201}
\preprint{MSU-HEP-041201}
\preprint{UH-511-1062-04}

\title{
\PRE{\vspace*{1.5in}}
Neutralino Cold Dark Matter in a One Parameter Extension\\ 
of the Minimal Supergravity Model
\PRE{\vspace*{0.3in}}
}

\author{Howard Baer}\author{Azar Mustafayev}\author{Stefano Profumo}
\affiliation{Dept. of Physics,
Florida State University, Tallahassee, FL 32306, USA
\PRE{\vspace*{.1in}}
}
\author{Alexander Belyaev}
\affiliation{Dept. of Physics and Astronomy,
Michigan State University, East Lansing, MI 48824, USA
\PRE{\vspace*{.1in}}
}
\author{Xerxes Tata}
\affiliation{Dept. of Physics and Astronomy,
University of Hawaii, Honolulu, HI 96822, USA
\PRE{\vspace*{.2in}}
}

%\date{November 24, 2004}

\begin{abstract}
\PRE{\vspace*{.1in}} 
Within the minimal supergravity model (mSUGRA) framework, the
expectation for the relic density of neutralinos exceeds the WMAP
determination, unless neutralinos {\it a})~have a significant higgsino
component, {\it b})~have a mass close to half that of a heavy Higgs
boson, or {\it c})~can efficiently co-annihilate with a charged or
colored particle. Within a 1-parameter extension of the mSUGRA model 
which includes non-universal Higgs masses, we
show that agreement with the WMAP data can be obtained over a wide range
of mSUGRA parameters for scenarios {\it a}) and {\it b}), so that the
phenomenological implications may be much more diverse than in mSUGRA.
We show that direct and/or indirect detection of neutralino dark matter
should be possible at various current and planned facilities.

\end{abstract}

%12.60.-i   Models beyond the standard model
%95.35.+d   Dark matter

\maketitle

Supersymmetry (SUSY) is a novel spacetime symmetry between bosons and fermions.
%Supersymmetric models of particle physics include a broken supersymmetry
%which yields superpartners of each Standard Model (SM) particle with mass
%expected in the range $\sim 100-1000$ GeV. 
In realistic models, SUSY is broken at the weak scale
implying that all Standard Model (SM) particles must have superpartners
with masses in the range $\sim 100-1000$ GeV that will be accessible to
colliders.
The lightest neutralino $\tz_1$ of $R$-parity conserving SUSY models 
is an especially attractive and well-motivated candidate 
for cold dark matter (CDM) in the universe\cite{goldberg}.  
The WMAP collaboration\cite{Spergel:2003cb} has
recently determined the relic density of CDM in the universe to be
$\Omega_{CDM}h^2=0.113\pm 0.009$ ($1\sigma$); this measurement,
especially its implied {\it upper limit}, provides a powerful constraint
for any model of particle physics that includes a candidate for CDM.

%In supersymmetric models of particle physics, 
The present neutralino relic density can be
determined by solving the Boltzmann equation for neutralinos in a
Friedmann-Robertson-Walker universe. The central aspect of the calculation
involves a computation of the thermally averaged neutralino
annihilation and co-annihilation cross sections.
Many analyses of SUSY CDM have been carried out within the  
paradigm minimal supergravity (mSUGRA)       
model\cite{msugra}. The free parameters of the mSUGRA model consist of
$m_0,\ m_{1/2},\ A_0,\ \tan\beta ,\ {\rm and}\ sign(\mu )$,
where $m_0$ is the common scalar mass, $m_{1/2}$ is the common
gaugino mass, and $A_0$ is a common trilinear soft SUSY breaking 
parameter all defined at the scale $Q=M_{GUT}\simeq 2\times 10^{16}$ GeV.
The parameter $\tan\beta$ is the ratio of weak scale Higgs field 
vacuum expectation values, and
$\mu$ is a superpotential parameter whose magnitude is constrained
by the requirement of radiative electroweak symmetry breaking (REWSB).
Once these model parameters are specified, then all
sparticle masses and mixings are determined, and scattering cross
sections, decay rates, relic density and dark matter detection rates 
may all be reliably calculated.

It has been increasingly recognized that
the typical value of $\Omega_{\tz_1}h^2$ as given by the mSUGRA model
significantly exceeds its WMAP upper limit.
Only specific regions of the mSUGRA model parameter space where
neutralinos can efficiently annihilate are in accord with the WMAP data.
These include: {\it i})~The bulk annihilation region at low values of
$m_0$ and $m_{1/2}$, where neutralino pair annihilation occurs at a
large rate via $t$-channel slepton exchange. {\it ii})~The stau (stop)
co-annihilation region at low $m_0$\cite{ellis_stau} (for special values
of $A_0$\cite{drees}) where $m_{\tz_1}\simeq m_{\ttau_1}$
($m_{\tz_1}\simeq m_{\tst_1}$) so that $\tz_1$s may co-annihilate with
$\ttau_1$s ($\tst_1$) in the early universe.  {\it iii})~The hyperbolic
branch/focus point (HB/FP) region at large $m_0\sim 3-7$ TeV (depending
sensitively on the assumed value of $m_t$) near the boundary of the
REWSB excluded region where $|\mu |$ becomes small, and the neutralinos
have a significant higgsino component, which facilitates annihilations
to $WW$ and $ZZ$ pairs\cite{ccn_fmm}.  {\it iv}). The $A$-annihilation
funnel, which occurs at very large $\tan\beta\sim 45-60$. In this case,
the value of $m_A\sim 2m_{\tz_1}$, so that neutralino annihilation in
the early universe is enhanced by the $A$ (and also $H$) $s$-channel
poles\cite{Afunnel}.
These allowed regions frequently occur either at edges
of parameter space, or for extreme values of mSUGRA parameters. In
this paper, we show that mSUGRA parameter choices that were previously 
disallowed by WMAP can in fact be brought into accord with the relic density
measurement via two possible solutions, in a well motivated one parameter
extension of the model. Thus, parameter choices thought to be irrelevant
for collider and other SUSY searches are in fact now allowed.
%Thus, what were previously considered as special situations for SUSY
%phenomenology are really quite generic.

The assumption of a universal soft SUSY breaking scalar mass $m_0$ is
phenomenologically motivated for matter scalars, since universality
guarantees in a super-GIM mechanism which suppresses dangerous flavor
changing neutral current processes\cite{georgi}.  Within the framework
of gravity-mediated SUSY breaking, universality for {\it all} scalars is
realized by assuming a flat K\"ahler potential, though non-universal
scalar masses are certainly possible\cite{soniweldon}.  While universal
masses for matter scalars is phenomenologically motivated, there is, no
motivation for assuming that the Higgs field scalars also have a common
GUT scale mass $m_0$. Indeed, in supersymmetric $SO(10)$ grand unified
theories, which are highly motivated by recent data on neutrino masses,
the matter multiplets of each generation belong to the 16 dimensional
spinorial representation of $SO(10)$, while the Higgs multiplets inhabit
a single 10 dimensional fundamental representation, and would have an
unrelated mass, assuming that the SUSY breaking mechanism is not
completely blind to gauge quantum numbers. In the simplest case, we may
naively expect the Higgs soft SUSY breaking squared masses to obey
$m_{H_u}^2=m_{H_d}^2\ne m_0^2$ so that this one parameter extension is
characterized by, \be m_0,\ m_\phi ,\ m_{1/2},\ A_0,\ \tan\beta ,\ {\rm
and}\ sign(\mu ), \ee where $m_\phi=
sign(m_{H_{u,d}}^2)\cdot\sqrt{|m_{H_{u,d}}^2|}$.\footnote{We stress that
we are using $SO(10)$ considerations only to guide our thinking, and do
not commit to any particular $SO(10)$ framework. As a result, many of
the usual consequences such as the unification of Yukawa couplings that
follow in the minimal $SO(10)$ model, no longer obtain in our
case. In particular, as we will see below, it is possible to break
electroweak symmetry radiatively even if the MSSM Higgs scalar doublets
have equal mass parameters at $Q=M_{GUT}$.}
We dub this the {\it non-universal} Higgs mass (NUHM1) model, to
distinguish it from the more general NUHM2 model\cite{an,bdqt} where
each Minimal Supersymmetric Standard Model (MSSM) Higgs scalar has an
independent mass parameter at $Q=M_{GUT}$.
The NUHM2 model has recently been
investigated by Ellis et al.\cite{ellis_nuhm}, and is motivated instead
by $SU(5)$ SUSY grand unification.  These NUHM2 models have also
been investigated recently within the context of Yukawa unified SUSY
models\cite{so10}.

%In this Letter, we report on the implications of the NUHM1 model,
%with emphasis on its predictions for neutralino cold dark matter.
We begin our analysis of the NUHM1 model 
by noting  that the parameter range for
$m_\phi$ need not be limited to positive values. 
Some authors have invoked the so-called GUT stability bound,
$sm_{\phi}^2+\mu^2(M_{GUT}) >0$ (with $s=sign(m_\phi )$), 
to avoid EWSB at too high a scale; see Ref.~\cite{gutstab} where the 
applicability of these bounds is discussed. Except to note
that these are hardly constraining for the cases that we present 
here, we do not consider them any further. 

We generate 
sparticle mass spectra using ISAJET v7.72\cite{isajet} upgraded to allow
the input of negative Higgs squared masses. ISAJET includes 
full one-loop radiative corrections
to all sparticle masses and Yukawa couplings, 
and minimizes the scalar potential using the
renormalization group improved effective potential 
renormalized at an optimized scale choice to account for leading two
loop terms. We evaluate the relic density using the IsaReD\cite{isared}
program, and to
evaluate the indirect CDM signals expected from the NUHM1
model, we adopt the DarkSUSY\cite{darksusy} package interfaced to
ISAJET.

\begin{figure}[tbp]
\postscript{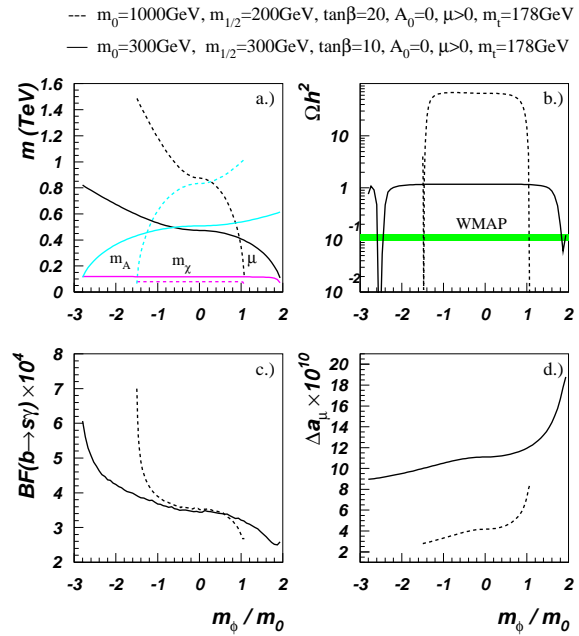}{0.95}
\caption{The values of {\it a}). $\mu$, $m_A$ and $m_{\tz_1}$ versus
$m_\phi /m_0$ for two cases in the NUHM1 model. We also show
{\it b}). $\Omega_{CDM}h^2$, {\it c}). $BF(b\to s\gamma )$ and 
{\it d}). $\Delta a_\mu$. The curves terminate on the left (right)
because $m_A^2<0$ ($\mu^2 <0$) so that REWSB is not obtained.
\label{fig:1}}
\end{figure}

As a first example, in Fig. \ref{fig:1} we show one of the 
critical aspects of the NUHM1 model, where we plot in {\it a}) the values of
$\mu$, $m_A$ and $m_{\tz_1}$ versus 
$m_\phi/m_0$, while
fixing other parameters as listed.
One can see that  $\mu$ becomes
much larger for $m_\phi\le -m_0$, and much smaller for $m_\phi > m_0$
compared to the case of   $m_\phi=m_0$.
The region of small $\mu$ is of particular interest since in
that case the lightest neutralino becomes more higgsino-like, and gives
rise to a relic density which can be in accord with
the WMAP determination\cite{drees_susy2004}. 
In the mSUGRA model, the higgsino-like $\tz_1$ region occurs in the HB/FP
region mentioned above, with $m_0\sim 3-7$ TeV.
We immediately see one important virtue of the NUHM1 model: the higgsino
annihilation region may be reached even with relatively low values of
$m_0$ and $m_{\phi}$. Thus, unlike the mSUGRA model case, 
the low $|\mu|$ region of the NUHM1 model can occur for small
values of $m_0$, $m_{\phi}$ and $m_{1/2}$, and so need not suffer from
fine tuning\cite{finetune}.

We also see from Fig. \ref{fig:1} that the value of $m_A$ can range
beyond its mSUGRA value for large values of $m_\phi$, to quite small values
when $m_\phi$ becomes less than zero\cite{javier}. In particular, when $m_A\sim
2m_{\tz_1}$, then neutralinos in the early universe may annihilate
efficiently through the $A$ and $H$ Higgs resonances, so that again
$\Omega_{CDM} h^2$ may be brought into accord with the WMAP
determination.  In the mSUGRA model, the $A$-annihilation funnel occurs
only at large $\tan\beta\gtrsim 45$ (55) for $\mu <0$ ($\mu >0$).  However,
in the NUHM1 model, the $A$-funnel region may be reached even for low
$\tan\beta$ values, if $m_\phi$ is taken to have negative values.  At low
$\tan\beta$, the $A$ and $H$ widths are much narrower than the large
$\tan\beta$ case.  We also show in frames {\it b}) the relic density
$\Omega h^2$, {\it c}) the branching fraction $BF(b\to s\gamma )$
(allowed range conservatively taken to be $\sim (2.6-4.5)\times 10^{-4}$), 
and {\it d}) the SUSY
contribution to the muon magnetic moment $\Delta a_\mu$.
We have also checked that the branching ratio $B(B_s \to \mu^+\mu^-) <
10^{-8}$ even for the smallest values of $m_A$ in Fig.~\ref{fig:1}.

To understand the behavior of the $\mu$ parameter and $m_A$ in the NUHM1
model, we define $\Delta m_{H_{u,d}}^2 \equiv m_{H_{u,d}}^2({\rm
  NUHM1})-m_{H_{u,d}}^2({\rm mSUGRA})$, and obtain the one loop
renormalization group equation 
\begin{eqnarray}
\frac{d\Delta m_{H_u}^2}{dt} &\simeq \frac{2}{16\pi^2}\times
3f_t^2[X_t({\rm NUHM1})-X_t({\rm mSUGRA})]\nonumber\\ 
&\simeq
\frac{3}{8\pi^2}f_t^2\Delta m_{H_u}^2,
\label{Dmhurge}
\end{eqnarray}
where $t=\log (Q)$, $f_{t}$ is the top quark Yukawa
coupling, and $X_t=m_{Q_3}^2+m_{\tst_R}^2+m_{H_u}^2+A_t^2$.
Eq.~(\ref{Dmhurge}) can be readily integrated to obtain, 
$$\Delta m_{H_u}^2(weak)=\Delta m_{H_u}^2(GUT) \times e^{-J_t}\;,$$
where $J_t=\frac{3}{8\pi^2}\int dt f_t^2 > 0$. We see that $\Delta
m_{H_u}^2$ maintains its sign under RG evolution, and furthermore,
reduces in magnitude in the evolution from high to low scales. The same
considerations also apply to $\Delta m_{H_d}^2$, except that the effect is
much smaller because for the modest values of $\tan\beta$ that we
consider, $f_{b,\tau}\ll f_t$ so that $J_b$ and $J_{\tau}$ are both
$\ll 1$ and $\Delta m_{H_d}^2(GUT) \simeq \Delta m_{H_d}^2(weak)$. 
From the tree level minimization condition for EWSB in the MSSM
\begin{displaymath}
\mu^2 =\frac{m_{H_d}^2-m_{H_u}^2\tan^2\beta}{(\tan^2\beta -1)}-
{M_Z^2\over 2} ,
\end{displaymath}
we see that 
for moderate to large values of $\tan\beta$ (as favored by LEP2 Higgs boson
mass constraints), and $|m_{H_u}^2|\gg M_Z^2$, $\mu^2\sim -m_{H_u}^2$. 
We thus see that a large value of $\mu^2$ is expected if $m_{\phi}$ is
large and negative. In contrast, for $m_\phi > m_0$, we expect a
smaller weak scale magnitude of $m_{H_u}^2$, and hence of $\mu^2$.

The tree level pseudoscalar Higgs mass $m_A$ is given by
\begin{displaymath}
m_A^2=m_{H_u}^2+m_{H_d}^2+ 2\mu^2 \simeq m_{H_d}^2-m_{H_u}^2,
\end{displaymath}
where the last equality holds in the approximation $\mu^2 \sim - m_{H_u}^2$.
For values of $\tan\beta$ such that $J_b$ and $J_{\tau}$ can be
neglected compared to $J_t$, it is easy to see that
$$m_A^2({\rm NUHM1})-m_A^2({\rm
  mSUGRA})=(sm_{\phi}^2-m_0^2)(1-e^{-J_t}) \; $$ 
thereby accounting for the $m_A$ dependence in Fig.~\ref{fig:1}a. 

Most other sparticle masses are relatively 
invariant to changes in $m_\phi$ (justifying our approximation in the
last step of (\ref{Dmhurge})). An obvious exception occurs for the
chargino $\tw_1$ and neutralino $\tz_{1,2}$ masses, 
which become small when $\mu\alt M_{1,2}$, 
and the $\tz_1$ becomes increasingly higgsino-like. The other exception
occurs for third generation squark masses. 
In this case, the
$(\tst_L,\tb_L )$ and $\tst_R$ running masses depend on $m_{\phi}$
via the $f_t^2 X_t$ terms in the RG equations, and can be somewhat
suppressed for large positive values of $m_{\phi}^2$. 
%. Thus, when $X_t$ is small
%(for $m_H\ll 0$), the top and bottom squarks receive little RG
%suppression,
%and when $m_H$ is large, so that $X_t$ is large, those soft masses should
%receive large suppression, and so we find lighter third generation squarks.

In order to estimate the dark matter detection rates for the NUHM1 model,
we adopt a cored halo model 
(the Burkert profile, \cite{burkert}) which has been tested against a 
large sample of rotation curves in spiral galaxies \cite{salucci-burkert}, 
and whose density profile reads
\begin{equation}
\rho_B(r)=\frac{\rho_B^0}{(1+r/a)\left(1+(r/a)^2\right)},
\end{equation} 
with $a=11.7\ {\rm kpc}$ and $\rho_B^0=0.34\ {\rm GeV}\ {\rm cm}^{-3}$.
The particular configuration we use has been found after implementing 
all available dynamical constraints and numerical simulation 
indications on the halo mass-concentration correlation \cite{piero}. 
The corresponding velocity distribution has been self-consistently 
computed (see Ref.~\cite{piero} for details), thus allowing for a 
reliable comparison among direct and indirect 
detection techniques \cite{Profumo:2004ty}. 
%Other consistent halo model choices would typically yield 
%larger detection rates\cite{howiepap}, particularly as far as 
%indirect detection is concerned: in this respect our results 
%can be regarded as conservative {\em lower limits}. 

\begin{figure*}[tb]
\epsfig{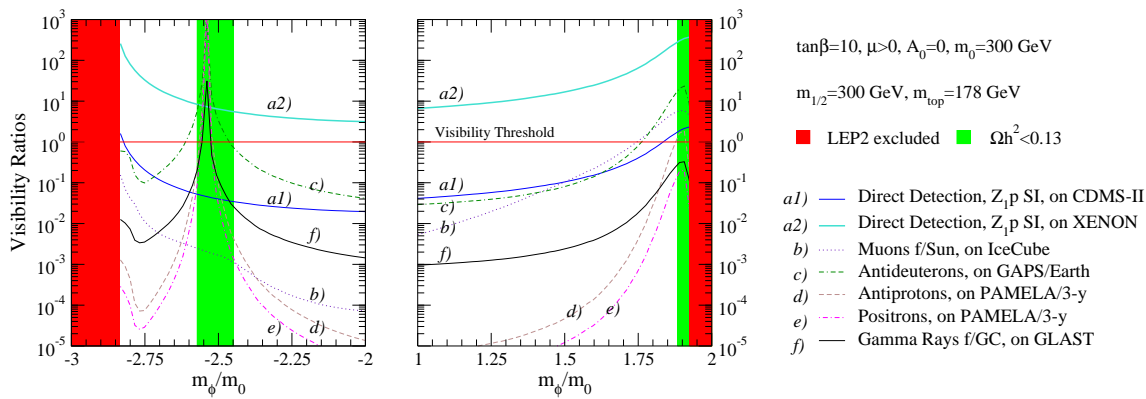}
\caption{Dark matter detection Visibility Ratios ({\em i.e.} 
signal-to-projected-sensitivity ratios) for various direct and 
indirect techniques, as a function of the GUT-scale non-universal 
Higgs mass parameter $m_\phi$. A Visibility Ratio larger than 1 implies 
that the model will be detectable. The dark (red) 
shadings to the left (right) 
of the plot indicate regions ruled out by the LEP2 bounds on 
$m_A$ ($m_{\tw_1}$). Within the lighter shaded
(green) regions 
a thermal neutralino relic abundance compatible with the WMAP 
upper bound \cite{Spergel:2003cb} is produced.
\label{fig:3}}
\end{figure*}

Following Ref.~\cite{Profumo:2004ty,Profumo:2004at}, 
we adopt here {\em Visibility Ratios} (VR), {\em i.e.} ratios of the 
expected signals from a given supersymmetric model over the 
projected future sensitivities in the particular detection technique, 
and for the corresponding neutralino mass. Direct detection refers to 
the spin-independent (SI) neutralino-proton scattering cross section, 
compared to the sensitivity of CDMS-II\cite{cdms} and of 
future ton-size experiments, 
such as XENON \cite{Aprile:2004ey}. 
We compute the flux of neutrino-induced muons produced by neutralino pair
annihilations in the sun, and compare
the flux against the projected sensitivity of the ${\rm km}^2$-size
detector IceCube (computed for an energy threshold of 1 GeV), 
taking into account the energy threshold mismatch and the
dependence of the detector sensitivity on the soft ({\em e.g.} $b\bar b$,
in the left panel of Fig.~\ref{fig:3}) and hard ({\em e.g.} $W^+W^-$, in the
right panel of Fig.~\ref{fig:3}) neutrino production channels, following
Ref.~\cite{esu}.
As regards antideuterons, we compute the solar-modulated flux, at solar
maximum, in the low energy range $0.1<T_{\overline{D}}<0.4$ GeV, where the
background is strongly suppressed \cite{dbar}. The VR is then defined as
the ratio of the computed $\overline D$ flux over the sensitivity of a GAPS
\cite{Mori:2001dv} detector placed on a satellite orbiting around the
Earth, and tuned to look for antideuterons in the mentioned very low
kinetic energy interval, corresponding to a $\overline D$ flux of 
$2.6\times 10^{-9} {\rm m}^{-2}{\rm sr}^{-1} {\rm GeV}^{-1} {\rm s}^{-1}$
\cite{Mori:2001dv}.
Regarding antiprotons and positrons, we adopt the quantity $I_\phi$, 
defined in Ref.~\cite{Profumo:2004ty}, and examine the sensitivity of the 
PAMELA experiment after 3 years of data-taking. 
Finally, the gamma ray flux from the galactic center (GC) refers to the 
sensitivity of the GLAST experiment to the integral photon flux 
above a 1 GeV threshold \cite{Morselli:2003xw}.

While only large values of $m_\phi$ will be probed via CDM detection at
CDMS-II, all the 
parameter space will be probed by future 
experiments such as XENON. Interestingly enough, 
we obtain observable rates for 
indirect detection experiments only in those regions where a 
neutralino relic abundance compatible with the WMAP upper 
bound \cite{Spergel:2003cb} is produced. 
The first cosmologically allowed parameter space region, 
lying in the range $-2.57<m_\phi/m_0<-2.45$, features a large neutralino 
annihilation cross section, due to the proximity of heavy Higgs 
resonances\cite{bo}. This yields detectable rates in all experiments
apart from IceCube, where the rate of neutralino capture inside the sun 
depends critically on the spin-dependent neutralino-nucleon 
coupling, which is suppressed in this large $|\mu|$ region:
%by a large value of the $\mu$ parameter: 
see Fig.~\ref{fig:3}. In the second viable parameter 
space region ($1.83<m_\phi/m_0<1.93$), a large higgsino fraction implies 
instead detectability at IceCube, as well as at future antiprotons 
and antideuterons searches. Though positrons and gamma rays rates 
are enhanced in this region, they will still be slightly below 
the respective planned experimental sensitivities. We note, however, that 
detection rates from halo annihilations are very sensitive to the
CDM halo profile which is assumed.
More quantitatively, we find that with an extremely cuspy profile (such as
the adiabatically contracted NFW profile of Ref.~\cite{piero}), the direct
detection and neutrino telescope rates increase only by around 20\%, the
$\overline D$ flux by $\sim 10^1$, positrons and
antiprotons VR's by $\sim 10^2$, and the gamma ray flux from the 
GC by more than $10^4$ (see also the discussions in Ref.~\cite{bo} and
\cite{Profumo:2004ty}).

{\it Summary:} 
The mSUGRA model prediction of the CDM relic density can 
be brought into accord with the WMAP measured value only if 
the neutralino annihilation cross section is enhanced, usually
due to $A$ and $H$ boson resonances, or by increasing
the higgsino content of the neutralino. Within the mSUGRA framework, these
possibilities occur only if $\tan\beta$ is very large or 
scalar masses are in the HB/FP region at very large $m_0$. 
We have shown that in the NUHM1 model, 
a well motivated 1-parameter extension of mSUGRA, neutralino 
annihilation via heavy Higgs resonances can occur even for
low values of $\tan\beta$, and neutralino annihilation via higgsino 
components can occur at low values of $m_0$. 
Thus, a much wider range of collider
signals and low energy phenomenology is possible compared to
expectations from the mSUGRA model.

This research was supported in part by grants from
the United States Department of Energy.

%For most choices of mSUGRA model parameters, the
%neutralino relic density may be brought into accord with the WMAP value
%by either {\it i}) decreasing $m_\phi$ to large negative values, where
%the $A$-resonance occurs, or {\it ii}) by increasing $m_\phi$ until the
%neutralino becomes sufficiently higgsino-like to yield again a relic
%density in accord with WMAP. It is well-known that this situation occurs
%in the NUHM2 model, where the parameters $m_{H_u}^2$ and $m_{H_d}^2$ may
%be traded for $m_A$ and $\mu$. That it also occurs in the $SO(10)$
%inspired NUHM1 model is gratifying in that many mSUGRA model parameter
%choices which were thought to be excluded are in fact allowed in this
%well-motivated one-parameter extension of the model.

%%%%%%%%%%%%%%%%%%%%%%%%%%%%%%%%%%%%%%%%%%%%%%%%%%%%%%

\end{document}